\documentclass[showpacs,preprintnumbers]{revtex4}
\usepackage{graphicx}
\usepackage{epsfig}
\usepackage{bm}

\makeatletter \leftline{\epsfbox{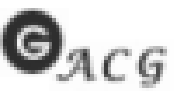}}
 \leftline
{\footnotesize{\textbf{\textsf{GACG-04/03}}}}
\begin{document}
\title{Quasinormal Modes of Extremal BTZ Black Hole}
\author{Juan Cris\'{o}stomo}
\altaffiliation{Permanent address: Departamento de F\'{i}sica,
Facultad de F\'{i}sica y Matem\'aticas, Universidad de
Concepci\'on. Casilla 160-C, Concepci\'on, Chile. e-mail:
jcrisost@gacg.cl}
\author{Samuel Lepe}
\altaffiliation{e-mail: slepe@gacg.cl}
\author{Joel Saavedra}
\altaffiliation{e-mail: joelium@gacg.cl}
\address{Instituto de F\'{\i}sica, Facultad de Ciencias B\'{a}sicas y Matem\'{a}ticas,
Pontificia Universidad Cat\'{o}lica de Valpara\'{\i}so, Avenida
Brasil 2950, Valpara\'{\i}so, Chile.}
\date{\today}

\begin{abstract}
Motivated by several pieces of evidence, in order  to show that extreme
black holes cannot be obtained as limits of non-extremal black
holes,  in this article we calculate explicitly
quasinormal modes for Ba\~{n}ados, Teitelboim and Zanelli (BTZ)
extremal black hole and we showed that the imaginary part of the
frequency is zero. We obtain exact result for the scalar an
fermionic perturbations. We also showed that the frequency is
bounded from below for the existence of the normal modes
(non-dissipative modes).
\end{abstract}
% PACS, the Physics and Astronomy
                             % Classification Scheme.
%\keywords{Suggested keywords}%Use showkeys class option if keyword
                              %display desired
\maketitle

%%%%%%%%%%%%%%%%%%%%%%%%%%%%%%%%%%%%%%%%%%%

\section{\label{sec:level1} Introduction}

%%%%%%%%%%%%%%%%%%%%%%%%%%%%%%%%%%%%%%%%%%%

There are several pieces of evidence to show that extremal Black
Holes (BH) cannot be obtained as limits of non-extremal BH. First
of all, these two classes are topologically different. For
example, in $2+1$ dimensions, the topology of an extremal BH is an
annulus, while the non-extremal BH is topologically like a
cylinder \cite{teitel}. Moreover, the extremal BH has zero Hawking
temperature ensuring stability against Hawking's radiation which
is not the case for the non-extremal BH. In the language of
thermodynamics, many authors have showed, at least
semiclassically, that the entropy of an extremal BH is zero
\cite{teitel, horowitz, wilczek, DAS} in contrast with the
non-extremal case where the entropy is proportional to the area.
In addition, the classical absorption cross section (or greybody
factor) for non extreme BH in three and four dimensions, turns out
to be proportional to the area of the horizon \cite{area} while in
Refs. \cite {gm,lmsv} it has been shown that the extremal BTZ BH
has null greybody factor for scalar and fermion particles. By
virtue of all these differences, it is clear that both systems
must be treated differently \cite{Wang:1999sk} and it is not
excepted to pass form one to the other by some procedure limits.

In the past few years there has been a growing interest in
studying the QNMs (quasinormal modes) spectrum of BH
\cite{qsnm,horowitz2} and its connection to conformal field theory
\cite{conformal1,qsnm21}. Unfortunately, finding an exact solution
of the QNMs is a very hard task and in some cases, it is almost
impossible to get an analytic expression. In spite of the
differences mentioned above, studies have focused on non-extremal
BH or near extremal BH, aside extremal case as limit of these last
ones. In this paper, we give another evidence based on QNMs ,
specifically that this limit procedure does not work in the case
of the extremal BTZ BH. Indeed, we will show that the QNMs of the
extremal BTZ BH do not exist for scalar and fermionic
perturbations. This conclusion does not occur in the non-extremal
case where the QNMs are known \cite{conformal1,qsnm212}. We
discuss the boundary conditions of the states at spatial infinity,
where there is no plane wave solution. The paper is organized as
follows : in the following section we review the Klein Gordon and
Dirac equations in a curved spacetime as well as the
$2+1$-dimensional extremal black hole. In the next section, we
study the scalar perturbations and their QNMs. Then, we repeat the
same computations for fermionic perturbations. The last section
contains our conclusions and remarks.

%%%%%%%%%%%%%%%%%%%%%%%%%%%%%%%%%%%%%%%%%%%%%%%%%%%%%%%%%%%%%%%%%%%%

\subsection{Klein Gordon and Dirac Equations in a Curved Space Time}

%%%%%%%%%%%%%%%%%%%%%%%%%%%%%%%%%%%%%%%%%%%%%%%%%%%%%%%%%%%%%%%%%%%%
In order to compute the QNMs we must solve the Klein Gordon as
well as the Dirac equations in the three-dimensional extremal
black hole background. In this section, we review these equations
in order to make the discussion self-contained.

The wave equation in the scalar case is given by
\begin{equation}
\bigg[ \frac{1}{\sqrt{-g}}\partial _{\mu }(\sqrt{-g}g^{\mu \nu
}\partial _{\nu })-m^{2}\bigg] \psi =0\mbox{,}  \label{kg}
\end{equation}
and, assuming spherical or axial symmetry, this equation can be
rewritten as
\begin{equation}
\frac{d^{2}R}{dr^{2}}+(k^{2}-V_{eff}(r))R=0\mbox{,}
\label{radial}
\end{equation}
where $V_{eff}$ is the effective potential produced by the BH
background \cite{birrel}.

For the fermionic perturbations, we need to solve the Dirac
equation given by \cite{naka}
\begin{equation}
\gamma ^{a}E_{a}^{\mu }\left( \partial _{\mu }-\frac{1}{8}\omega
_{bc\mu }\left[ \gamma ^{b},\gamma ^{c}\right] \right) \Psi =m\Psi
\mbox{,} \label{DE}
\end{equation}
where $E_{a}^{\mu }$ is the inverse triad which satisfies
$E_{a}^{\mu }e_{\mu }^{b}=\delta _{a}^{b}$ and $\omega _{\mu }$ is
the spin-connection. The set of matrices $\left\{ \gamma
^{a}\right\} $ are the Dirac matrices in the tangent space which
satisfy the Clifford algebra
\begin{equation}
\left\{ \gamma ^{a},\gamma ^{b}\right\} =2\eta ^{ab}\mbox{.}
\label{cliff}
\end{equation}

\textbf{\emph{Extremal $2+1$ black hole metric}}. In a
$2+1$-dimensional
spacetime, the Einstein equations with negative cosmological constant $%
\Lambda =-\ell ^{-2}$ have the following solution \cite{btz}
\begin{equation}
ds^{2}=-N^{2}\left( r\right) dt^{2}+N^{-2}\left( r\right)
dr^{2}+r^{2}\left[ d\phi +N^{\phi }\left( r\right) dt\right]
^{2}\mbox{,}  \label{btz}
\end{equation}
where the lapse $N^{2}(r)$ and shift $N^{\phi }(r)$ functions are
given by
\begin{eqnarray}
N^{2}\left( r\right) &=&-M+\frac{r^{2}}{\ell ^{2}}+\frac{J^{2}}{4r^{2}}, \\
N^{\phi }\left( r\right) &=&-\frac{J}{2r^{2}}\mbox{.}
\end{eqnarray}
Here $M$ and $J$ are the mass and the angular momentum of the
black hole, respectively.

The lapse function vanishes when
\begin{equation}
r_{\pm }=r_{ex}\left[ 1\pm \sqrt{1-\frac{J^{2}}{M^{2}\ell ^{2}}}\right] ^{%
\frac{1}{2}}\mbox{,}
\end{equation}
and therefore, the solution (\ref{btz}) is defined for $r_{+}<r<\infty $, $%
-\pi <\phi <\pi $ and $-\infty <t<\infty $. The extremal solution
corresponds to $J^{2}=M^{2}\ell ^{2}$, which implies that $r_{\pm
}=r_{ex}=\ell \sqrt{M/2}$. Hence the line element (\ref{btz})
becomes
\begin{equation}
ds_{ex}^{2}=-\left( \frac{r^{2}}{\ell ^{2}}-2\frac{r_{ex}^{2}}{\ell ^{2}}%
\right) dt^{2}+\frac{\ell ^{2}r^{2}}{\left( r^{2}-r_{ex}^{2}\right) ^{2}}%
dr^{2}-2\frac{r_{ex}^{2}}{\ell }dtd\phi +r^{2}d\phi ^{2}\mbox{.}
\label{btze}
\end{equation}

Let us now study the scalar perturbations for this geometry.

%%%%%%%%%%%%%%%%%%%%%%%%%%%%%%%%%%%%%%%%%%%

\section{Scalar Perturbations}

%%%%%%%%%%%%%%%%%%%%%%%%%%%%%%%%%%%%%%%%%%%

Equation (\ref{kg}), in the above metric (\ref{btze}) can be
solved considering the following Ansatz
\begin{equation}
\psi =\psi _{0}\,e^{i\omega t}e^{in\phi }\,R(r)\mbox{,}
\label{ans}
\end{equation}
where $R(r)$ is an unknown function and $\psi _{0}$ is a constant. Plugging (%
\ref{ans}) into the equation (\ref{kg}), the radial function
$R(r)$ satisfies
\begin{eqnarray}
&&\biggl(\frac{(r^{2}-r_{ex}^{2})^{4}}{r^{2}l^{4}}\biggr)R^{\prime
\prime
}(r)+\biggl(\frac{(r^{2}-r_{ex}^{2})^{3}}{r^{3}l^{4}}\biggr)%
(3r^{2}-r_{ex}^{2})R^{\prime }(r)+  \nonumber \\
&&[(\omega l+n)((\omega l-n)r^{2}+2nr_{ex}^{2})-\frac{m^{2}}{l^{2}}%
(r^{2}-r_{ex}^{2})^{2}]R(r)=0\mbox{.}  \label{kg2} \\
&&  \nonumber
\end{eqnarray}
The solution of this equation is given as a linear combination
\cite{gm}
\begin{equation}
R(r)=AR^{(1)}(r)+BR^{(2)}(r)\mbox{,}  \label{rr}
\end{equation}
where the functions $R^{(1)}(r)$ and $R^{(2)}(r)$ read
\begin{eqnarray}
R^{(1)}(r) &=&e^{-i\Omega
_{+}\frac{r_{ex}^{2}}{r^{2}-r_{ex}^{2}}}\left(
\frac{r_{ex}^{2}}{r^{2}-r_{ex}^{2}}\right) ^{s_{+}}F[s_{+}+i\frac{\Omega _{-}%
}{2},2s_{+},2i\Omega _{+}\frac{r_{ex}^{2}}{r^{2}-r_{ex}^{2}}],
\label{sol1}
\\
R^{(2)}(r) &=&e^{-i\Omega
_{+}\frac{r_{ex}^{2}}{r^{2}-r_{ex}^{2}}}\left(
\frac{r_{ex}^{2}}{r^{2}-r_{ex}^{2}}\right) ^{s_{-}}F[s_{-}+i\frac{\Omega _{-}%
}{2},2s_{-},2i\Omega
_{+}\frac{r_{ex}^{2}}{r^{2}-r_{ex}^{2}}]\mbox{.}
\nonumber \\
&&  \label{sol2}
\end{eqnarray}
Here $s_{\pm }=\frac{1}{2}(1\pm \sqrt{1+m^{2}}),$ $\Omega _{\pm }=\frac{1}{%
\sqrt{2M}}(\omega \pm n)$ (we have put $l=1$) and $F[a,c,z]$ is
the confluent hypergeometric function (known as Kummer's
solution).

As the BTZ BH is asymptotically Anti de Sitter (AdS), the
definition of the QNMs is different from the one used in an
asymptotically flat space \cite{Roman} \cite{Wang}. In the former
case, this problem was well described in Ref. \cite{horowitz2}
where they defined QNMs as solutions which are purely ingoing at
the horizon, and vanishing at infinity. The vanishing boundary
condition implies that the constant $B$ in (\ref{rr}) must be zero
and in the asymptotic limit the wave function becomes
\begin{equation}
\psi _{\infty }=Ae^{i\omega t}e^{in\phi }e^{-i\Omega _{+}\left( \frac{%
r_{ex}^{2}}{r^{2}-r_{ex}^{2}}\right) }\left( \frac{r_{ex}^{2}}{%
r^{2}-r_{ex}^{2}}\right) ^{s_{+}}\mbox{.}  \label{qsnsc1}
\end{equation}
It is simple to see that the required boundary condition is
automatically satisfied, showing therefore the absence of QNMs.
This result is in contrast with the non-extremal case where QNMs
are proportional to the quantized imaginary part of the frequency
\cite{conformal1}. Moreover, this result is in agreement with the
null absorption cross section for the scalar case \cite {gm}.
Now, we turn to confirm again this result by showing that the
condition of the vanishing flux at infinity is satisfied. For
simplicity, we consider the coordinate
$z=\frac{r_{ex}^{2}}{r^{2}-r_{ex}^{2}}$ for which the conserved
radial current becomes

\begin{equation}
J_{r}(z)=R^{*}(z)\frac{d}{dz}R(z)-R(z)\frac{d}{dz}R^{*}(z).
\label{scalarcurrent}
\end{equation}
 Due to the regularity condition at infinity, the only contribution to the
current $J_{r}(z)$ comes from $R^{(1)}(r)$. Then, the conserved
current is expressed as
\begin{equation}
J_{r}(z)=-i\left| A\right| ^{2}\Omega
_{+}z^{2s_{+}}F[s_{+}+i\frac{\Omega
_{-}}{2},2s_{+},2i\Omega _{+}z]F[s_{+}-i\frac{\Omega _{-}}{2}%
,2s_{+},-2i\Omega _{+}z],  \label{scalarcurrent2}
\end{equation}
and the flux is given by

\begin{equation}
\mathcal{F}=\sqrt{g}\frac{1}{2i}J_{r}(z). \label{flux1}
\end{equation}
It is straightforward to see that the flux vanishes as $z$ goes
to $0$ which confirms the absence of QNMs for extremal BTZ BH
under scalar perturbations. We also point out that this result is
independent of the value of $m$.

\section{Fermionic Perturbations}

In order to solve Dirac's equation (\ref{DE}) with the extremal
BTZ metric, it is convenient to define a dimensionless set of
coordinates $\left\{ u,v,\rho \right\} $ as follows (with $\ell $
restored)
\begin{equation}
u=\frac{t}{\ell }+\phi
,\,\,\,\,\,\,\,\,\,\,\,\,\,\,\,\,\,\,\,\,\,\,\,\,\,\,\,v=\frac{t}{\ell }%
-\phi ,\,\,\,\,\,\,\,\,\,\,\,\,\,\,\,\,\,\,\,\,\,\,\,\,\,\,\,e^{2\rho }=%
\frac{r^{2}-r_{ex}^{2}}{\ell ^{2}}\mbox{,}
\end{equation}
where $u,v$ and $\rho $ range from $-\infty $ to $\infty $. In the space $%
\{u\times v\}$, two points $(u_{1},v_{1})$ and $(u_{2},v_{2})$ are
identified if they satisfy $u_{1}=v_{2}$ and $v_{1}=u_{2}$, for
any value of $\rho $.

Choosing the Dirac matrices to be given by
\begin{equation}
\gamma ^{1}=-i\sigma
^{3},\,\,\,\,\,\,\,\,\,\,\,\,\,\,\,\,\,\,\,\,\,\,\,\,\,\,\,\,\,\,\gamma
^{2}=\sigma
^{1},\,\,\,\,\,\,\,\,\,\,\,\,\,\,\,\,\,\,\,\,\,\,\,\,\,\,\,\gamma
^{3}=\sigma ^{2}\mbox{,}
\end{equation}
where $\sigma ^{i}$ are the Pauli matrices and, taking the
following Ansatz for the wave function
\[
\Psi (u,v,\rho )=\left(
\begin{array}{c}
{\mathcal{U}}(u,v,\rho ) \\
{\mathcal{V}}(u,v,\rho )
\end{array}
\right) \mbox{,}
\]
the Dirac equation becomes
\begin{eqnarray}
\left[ -i\left( \frac{2r_{ex}e^{-2\rho }}{\ell ^{2}}\partial _{u}+\frac{%
\partial _{v}}{r_{ex}}\right) -(\frac{1}{2\ell }+m)\right] {\mathcal{U}}%
+\left[ \frac{\partial _{v}}{r_{ex}}-\frac{i}{\ell }\left(
\partial _{\rho
}-1\right) \right] \mathcal{V} &=&0,  \label{DEC1} \\
\left[ \frac{\partial _{v}}{r_{ex}}+\frac{i}{\ell }\left( \partial
_{\rho
}-1\right) \right] {\mathcal{U}}+\left[ i\left( \frac{2r_{ex}e^{-2\rho }}{%
\ell ^{2}}\partial _{u}+\frac{\partial _{v}}{r_{ex}}\right)
-(\frac{1}{2\ell }+m)\right] {\mathcal{V}} &=&0\mbox{.}
\label{DEC2}
\end{eqnarray}
Let us now look for solutions of this equation of the form
\begin{equation}
\Psi \left( u,v,\rho \right) =e^{i\left( \alpha u+\beta v\right)
}\left(
\begin{array}{c}
F(\rho ) \\
G(\rho )
\end{array}
\right) \mbox{,}  \label{ansatz}
\end{equation}
where $\alpha $ and $\beta $ are two constants related to the
angular and
temporal eigenvalues of the solution of (\ref{DE}) in the coordinates $%
\left\{ t,\phi ,r\right\} $. This means that if the solution behaves like $%
e^{i(n\phi +\omega t)}$, then
\begin{equation}
\alpha =\frac{1}{2}\left( \omega \ell +n\right)
,\,\,\,\,\,\,\,\,\,\,\,\,\,\,\,\,\,\,\,\,\,\,\,\,\,\,\,\,\,\,\,\,\,\,\,\,%
\beta =\frac{1}{2}\left( \omega \ell -n\right) \mbox{.}
\end{equation}
For the $\rho $-dependent part of the equation it is useful to define $%
z=e^{-2\rho }$ and thus, the functions $F(z)$ and $G(z)$ satisfy
the following system
\begin{eqnarray}
\left[ \frac{2\alpha r_{ex}}{\ell ^{2}}z+\frac{\beta }{r_{ex}}-\frac{1}{%
2\ell }-m\right] F\left( z\right) +i\left[ \frac{\beta }{r_{ex}}+\frac{1}{%
\ell }\left( 2z\frac{d}{dz}+1\right) \right] G\left( z\right)
&=&0\mbox{,}
\label{ED1} \\
i\left[ \frac{\beta }{r_{ex}}-\frac{1}{\ell }\left(
2z\frac{d}{dz}+1\right)
\right] F\left( z\right) -\left[ 2\frac{\alpha r_{ex}}{\ell ^{2}}z+\frac{%
\beta }{r_{ex}}+\frac{1}{2\ell }+m\right] G\left( z\right)
&=&0\mbox{.} \label{ED2}
\end{eqnarray}
This system of first order coupled differential equations can be
transformed into a second order system. Indeed, using the equation
(\ref{ED1}) to express $F(z)$ as a function of $G(z)$ and its
first derivative and then defining a new variable $x=\alpha
(\,r_{ex}/\ell )\,z$, one finds that $G(x)$ satisfies the
following second order equation
\begin{equation}
A(x)G^{\prime \prime }(x)+B(x)G^{\prime }(x)+C(x)G(x)=0\mbox{.}
\label{eqg}
\end{equation}
The function $A,B$ and $C$ are given by
\begin{eqnarray}
A(x) &=&(\delta -x)x^{2}\mbox{,} \\
B(x) &=&(2\delta -x)x\mbox{,} \\
C(x) &=&-x^{3}+(\delta -\tilde{\beta})\,x^{2}+\frac{1}{4}\bigg[(\tilde{\beta}%
+1)^{2}+4\delta (2\tilde{\beta}+\delta )\bigg]x-\frac{\delta
}{4}\left( (2\delta +\tilde{\beta})^{2}-1\right) \mbox{,}
\end{eqnarray}
where the constants $\delta $ and $\tilde{\beta}$ are
\begin{equation}
\delta =\frac{1}{2}\left( \ell \,m_{\mbox{\scriptsize{eff}}}-\tilde{\beta}%
\right)
,\,\,\,\,\,\,\,\,\,\,\,\,\,\,\,\,\,\,\tilde{\beta}=\frac{\ell
\beta }{r_{ex}}\mbox{.}
\end{equation}
and the effective mass is
$m_{\mbox{\scriptsize{eff}}}=m+\frac{1}{2\ell }$. The same
procedure can be done for the function $F(\rho )$ which yields to
a similar result (for details, see Ref. \cite{lmsv}). In order to
solve the equation expressed in (\ref{eqg}), we need to
distinguish between two cases according to a vanishing or
non-vanishing delta.

$\delta =0$. For a vanishing $\delta $ (i. e. $\beta =r_{\mbox
ex}(m+1/2l)$, the equation (\ref{eqg}) reduces to
\begin{equation}
x^{2}G^{\prime \prime }(x)+xG^{\prime }(x)+(x^{2}+\tilde{\beta}x-\frac{1}{4}(%
\tilde{\beta}+1)^{2})G(x)=0\mbox{,}
\end{equation}
whose solution is given by
\begin{equation}
G(x)=e^{-ix}\left( Px^{s}F\left[ s+1/2+i(s-1/2),1+2s,2ix\right]
+Qx^{-s}F\left[ -s+1/2+i(s-1/2),1-2s,2ix\right] \right) \mbox{.}
\end{equation}
Here $s=(\tilde{\beta}+1)/2$, $F[c,d;x]$ is the confluent
hypergeometric function, with $P$ and $Q$ are two complex
constants.

For small $x$, the solution $G(x)$ behaves like
\begin{equation}
G(x)\sim e^{-ik}(Px^{s}+Qx^{-s})\mbox{,}  \label{qsnmf}
\end{equation}
and hence the vanishing boundary condition at infinity ($\rho
\rightarrow \infty $ i. e. $x\rightarrow 0$) implies that $Q$ must
be zero. Thus, as we found in the scalar case, we have proved the
absence of the QNMs for the fermionic case. We want to note that
the absence of QNMs is different from the non-extremal case where
the QNMs are proportional to the quantized imaginary part of the
frequency \cite{conformal1}.

Another interesting feature of the solution $G$ (with $Q=0$) is that, for $%
x\approx \infty $, $G$ behaves like
\begin{equation}
G(x)\sim P\frac{e^{\pm
i\tilde{\beta}ln(x)/2}}{\sqrt{x}}\rightarrow 0\mbox{.}
\end{equation}
This also shows that the function $G(x)$ is vanishing at the
horizon, in agreement with a null absorption cross section for the
fermionic fields in an extremal BTZ background \cite{lmsv}.

$\delta \neq 0$. For a non-vanishing $\delta $, it is more
convenient to define a new variable $y=x/\delta $ and a function
$G(y)$ as
\begin{equation}
G(y)=\frac{1-y}{\sqrt{y}}D(y)\mbox{.}
\end{equation}
In terms of $D(y)$, the equation (\ref{eqg}) turns out to be
\begin{eqnarray}
y^{2}(1-y)^{2}D^{\prime \prime }+y(1-y)(1-2y)D^{\prime
}+\bigg[\delta
^{2}y^{4}+\delta \left( \tilde{\beta}-2\delta \right) y^{3} &-&\frac{\tilde{%
\beta}}{4}\left( 2+\tilde{\beta}+12\delta \right) y^{2}+  \nonumber \\
\frac{1}{4}\left( 5(\tilde{\beta}+\delta
)^{2}+2\tilde{\beta}(1+\delta )-4\right)
y-\frac{1}{4}(2\tilde{\beta}+\delta )^{2}\bigg]D &=&0\mbox{.}
\label{grosa}
\end{eqnarray}
As in the previous cases, we need to study the asymptotic behavior
of the solution of this equation to explore the possibility of
obtaining QNMs. Keeping terms up to first order, this limit
$y\rightarrow 0$ yields to
\begin{equation}
y^{2}D_{0}^{\prime \prime }+(1-y)yD_{0}^{\prime }+\left( -\frac{1}{4}(2%
\tilde{\beta}+\delta )^{2}+\frac{1}{4}\left( \tilde{\beta}(2+4\delta )-3(%
\tilde{\beta}^{2}-\delta ^{2}-4)\right) y\right) D_{0}=0\mbox{,}
\end{equation}
whose solution is
\begin{equation}
D_{0}(y)=P_{0}\,\,\,y^{-\frac{b}{2}}F\left[ \frac{1}{2}\left(
a+b\right) ,1-b,y\right] +Q_{0}\,\,\,y^{\frac{b}{2}}F\left[
\frac{1}{2}\left( a-b\right) ,1+b,y\right] \mbox{.}  \label{sol0}
\end{equation}
Here $P_{0}$ and $Q_{0}$ are two complex constants, $a=2+\frac{3}{2}(\tilde{%
\beta}^{2}-\delta ^{2})-\tilde{\beta}(1+2\delta )$ and $b=2\tilde{\beta}%
+\delta $ and $F[c,d;x]$ is the confluent hypergeometric function.
Finally the solution reads,
\begin{equation}
G_{0}(y)=P_{0}\,\,\,y^{-\frac{b+1}{2}}+Q_{0}\,\,y^{\frac{b-1}{2}}\mbox{.}
\label{juancho1}
\end{equation}
In order to satisfy the adequate boundary condition
\cite{horowitz2}, we require $P_{0}=0$ and $b>1.$ With these two
conditions, the vanishing boundary condition at infinity again is
automatically satisfied by (\ref {juancho1}) and this fact clearly
shows the absence of QNMs in this case.

On the other hand, it is straightforward to prove that the wave
function vanishes at horizon. Therefore the fields behavior is as
particles confined in a box. Allowing the presence of the normal
modes of vibration (non-dissipative modes).

At this point,let us remarking that the condition $b>1$ implies
that the frequency is bounded from below for the existence of the
normal modes, i. e.
\begin{equation}
\frac{\omega }{\Omega
_{ex}}>n+\sqrt{\frac{M}{2}}-\frac{2}{3}mr_{ex}\mbox{,}
\label{cutfrecuencie}
\end{equation}
where $\Omega _{ex}=l^{-1}$is the angular velocity of the extremal
black hole. This result can be interpreted as that the fundamental
frequency is bounded by below. In order to excite the normal modes
of the space, frequency should be more bigger this limit. Then in
this case extreme BTZ BH behavior as non-dissipative system. On
the other hand, we conjectured that this inferior limit is related
with the quantization of the black hole horizon area.
\cite{setare}

However, as noted in Ref. \cite{conformal1}, the Dirichlet
condition is not adequate for the BTZ BH for some values of the
mass parameter. The authors in Ref. \cite{conformal1} consider
this fact as another motivation for imposing the vanishing flux
at infinity rather than the Dirichlet condition for
asymptotically AdS space-time. Then in the case of Dirac modes,
one can also impose vanishing flux at infinity (generally much
weaker than the Dirichlet condition), see Refs. \cite{conformal1}
and \cite{qsnm21}.
 From Eqs.  (\ref{ansatz}) and (\ref{ED1}) we obtain that

\begin{equation}
j^{\rho }(x)=\mathcal{A}\left( \frac{x}{\delta -x}\right) \Re \mbox{e}%
\left\{ i\,G(x)\frac{d}{dx}G^{*}(x)\right\} ,  \label{alfincarajo}
\end{equation}
where $\mathcal{A}$ is a constant and
\begin{equation}
\mathcal{F}=\mathcal{A}\,\sqrt{r_{ex}^{2}+\frac{\alpha \ell r_{ex}}{x}}%
\left( \frac{x}{\delta -x}\right) \Re \mbox{e}\left\{ i\,G(x)\frac{d}{dx}%
G^{*}\left( x\right) \right\} .  \label{otra}
\end{equation}

In the present case, $\delta =0$, and so the function $G(x)$ is
given by

\begin{equation}
G(x)=e^{-ix}\left( Px^{s}F\left[ s+1/2+i(s-1/2),1+2s,2ix\right]
+Qx^{-s}F\left[ -s+1/2+i(s-1/2),1-2s,2ix\right] \right) \mbox{,}
\end{equation}

while the flux reads
\begin{equation}
\mathcal{F}=-\mathcal{A}\sqrt{r_{ex}^{2}+\frac{\alpha \ell r_{ex}}{x}}%
x^{2s}F\left[ s+1/2+i(s-1/2),1+2s,2ix\right] F\left[
s+1/2-i(s-1/2),1+2s,-2ix\right].   \label{fluxfermion}
\end{equation}

In the limit $x$ goes to $0$, one has

\[
\mathcal{F}\rightarrow x^{2s-\frac{1}{2}},
\]
and thus, in order to satisfy the appropriate boundary condition,
we require that $Q=0 $ together with $2s-\frac{1}{2}>0$. Under
these two conditions, the absence of QNMs is ensured.

 Let us end this section with a remark concerning the frequency. Indeed, the
condition $2s-\frac{1}{2}>0$ again implies that the frequency is
bounded from below for the existence of the normal modes, i. e.

\[
\omega >\frac{n}{l}-\frac{r_{ex}}{l^{2}}
\]

For $\delta \neq 0$, it is possible to prove that the condition
(\ref{cutfrecuencie}) is also satisfied.

\section{Conclusion and Remarks}

Through of the exact solutions found in Refs. \cite{gm} and
\cite{lmsv} for the Klein Gordon and the Dirac equations in an
extremal BTZ background, we have explicitly discussed the absence
of QNMs in BTZ black holes.

In the scalar and fermionic cases we have shown that the vanishing
boundary conditions at infinity are automatically satisfied for
the exact solutions. This fact implies the absence of the QNMs in
the extremal BTZ BH against the non-extremal cases. Which is
agrees with several pieces of evidence that show that extremal BH
cannot be obtained as limits of non-extremal BH. In order to apply
adequate boundary condition, consistent with the presence of QNMs
in the AdS space, we have obtained an inferior limit for the
existence of normal modes. This last result is remarkable because
in general the perturbation equation provides only information
about the QNMs. We can observe that the extremal BTZ BH, behaves
as a non-dissipative system.

By other way, according to the AdS/CFT correspondence, the black
holes correspond to thermal states in the conformal field theory.
Due to the fact that the extremal black hole have a zero Hawking
temperature, it is possible to compute the zero temperature
2-point functions in an unique form (up to a normalization) from
conformal field theory \cite{Son:2002sd} \cite{Nunez:2003eq}
\cite{Abdalla:2002rm}. In this case, the retarded Green's function
will generally have poles and cuts on the real axis, corresponding
to stable states and multi-particles, respectively. On the other
hand, there also can be poles in the lower half $\omega $-plane.
The distance of the poles from the real line then determines the
decay time of such a resonance, this being relating with the QNM's
\cite{qsnm21}. In our case, the extremal black hole correspond to
a CFT in a cylinder of $(1+1)$ dimensions at zero temperature. The
imaginary pole part is null, showing that the extremal BTZ black
hole does not exhibit QNM's. On the other way, the real pole part
is different from zero and is quantized, showing the presence of
normal modes.

Then, from the CFT point of view, we can obtain another piece of
evidence about the distinction between the extremal and the
non-extremal cases, related to the different topologies (annulus
and cylinder respectively). On the other hand, the
AdS$_{3}$/CFT$_{2}$ correspondence contains information about
normal and quasinormal modes, and in the extremal case it is
possible to show the absence of QNM's according to the
perturbation equations developed in this article.

Finally, we hypothesize that these results are related with the
interpretation of the extremal black holes as fundamental
particles \cite {wilczek}.

\section{Acknowledgments}
We are grateful to A.Anabalon, M. Hassa\"{i}ne, R. Troncoso and
J. Zanelli for enlightening discussions. SL and JS are supported
by COMISION NACIONAL DE CIENCIAS Y TECNOLOGIA through FONDECYT
Grant N° 1040229 (SL) and Postdoctoral Grant 3030025 (JS). This
work is also partially supported by PUCV Grant No. 123.769/03
(SL). JC is supported by the Ministerio de Educaci\'{o}n through
a MECESUP Grant, FSM 9901. SL and JS thank the organizers of the
CECS Summer Meeting on Theoretical Physics for their kind
hospitality. We thank Dr. U. Raff for a careful reading of the
manuscript.

\end{document}